\documentclass[runningheads]{llncs}
\usepackage[T1]{fontenc}
\usepackage{graphicx}
\usepackage{amsmath}
\usepackage{amssymb}
\usepackage{booktabs}
\usepackage{url}
\usepackage{float}
\usepackage{xcolor}
\usepackage{tikz}
\usetikzlibrary{arrows.meta,calc,fit,positioning}
\begin{document}
%
\title{PyDoseRT Photon: Physics-Guided Pencil-Beam Dose Calculation with Neural Priors and Residual Correction for CT and MRI}
\titlerunning{PyDoseRT Photon}
\author{Attila Simk\'o\inst{1} \and
Lukas Zimmermann\inst{2,3} \and
Hermann Fuchs\inst{2,3} \and
Gerd Heilemann\inst{2,3}}
\authorrunning{A. Simk\'o et al.}
\institute{Department of Diagnostics and Intervention, Ume\aa{} University, Ume\aa, Sweden \and
Department of Radiation Oncology, Medical University of Vienna,
Vienna, Austria \and
Christian Doppler Laboratory for Image and Knowledge Driven Precision
Radiation Oncology, Medical University of Vienna, Vienna, Austria}
\maketitle
\begin{abstract}
We present a hybrid, physics-based analytical
pencil-beam (PB) dose engine, augmented by two small frozen neural physics
priors, followed by a 3-D convolutional residual-correction network (U-Net).

We address the DoseRAD2026\footnote{\url{https://doserad2026.grand-challenge.org/}} photon dose-prediction task with \emph{PyDoseRT
Photon}, a GPU PB engine implemented in PyTorch and corrected toward Monte
Carlo (MC) accuracy in three learned stages of decreasing physical
specificity. The engine reproduces the challenge's head-less MC source
exactly where it can and models the patient with a beam-quality-indexed
pencil kernel evaluated at the field's fluence-weighted radiological depth,
and TERMA source scaling at the interaction site. Two tiny neural priors
are trained through the frozen engine and then frozen themselves: a
39k-parameter 2-D fluence correction and a 48-parameter lateral
heterogeneity correction mixing mass-conserving Gaussian redistribution
operators. A compact 3-D U-Net (1.36M parameters) with a sequential
refinement branch then predicts, per control point (CP) in the
beam's-eye-view (BEV) frame, a bounded multiplicative gain and additive
residual from seven channels. All learned stages are
zero-initialized, so training starts from the analytical solution. For
MRI, an nnU-Net regression model synthesizes a CT that enters the identical
pipeline, with consecutively, the same trained corrector as the CT track.
Design choices were driven by the challenge ranking, in which runtime carries
double weight. The submitted method evaluated on our local CT and MR validation dataset achieved CP
MAE $0.0086$ and $0.0098$, IDD distance $0.0011$ and $0.0013$, plan MAE $0.0024$ and $0.0047$, gamma pass rate
(1\%/1\,mm) $99.12\%$ and $96.95\%$, with runtimes of $46\,\mathrm{s}$ and $49\,\mathrm{s}$, respectively.
\end{abstract}
\section{Introduction}
Volumetric-modulated arc therapy delivers dose through a sequence of small,
irregular apertures per arc, so the plan dose is the sum of many low-dose
segments whose individual accuracy is rarely scrutinized. Monte Carlo (MC)
transport is the accuracy reference but is too slow for iterative planning;
analytical pencil-beam (PB) algorithms~\cite{ref_ahnesjo} are fast but
degrade in lung, where lateral electron transport dominates, and at
bone/tissue interfaces. Recent work showed that a \emph{differentiable} dose
engine enables gradient-based plan optimization end-to-end~\cite{ref_pydose}.

We build on that engine for the DoseRAD2026 photon task. Two \emph{tiny}
neural modules were placed at physically meaningful points inside the
engine---on the source-plane fluence and on lateral energy transport---and
trained by back-propagating the challenge objective through the frozen
physics; once trained they were frozen and became part of the engine. A
compact volumetric U-Net then corrected what remains, per control point.
Every learned stage was zero-initialized and bounded, so the system degrades
gracefully to the analytical solution. Because the challenge ranks runtime
with double weight, all design decisions were taken on estimated leaderboard
position, not accuracy alone.

\section{Methods}
\subsection{Data}
\label{sec:data}
\begin{figure}[t!]
\centering
\includegraphics[width=\textwidth]{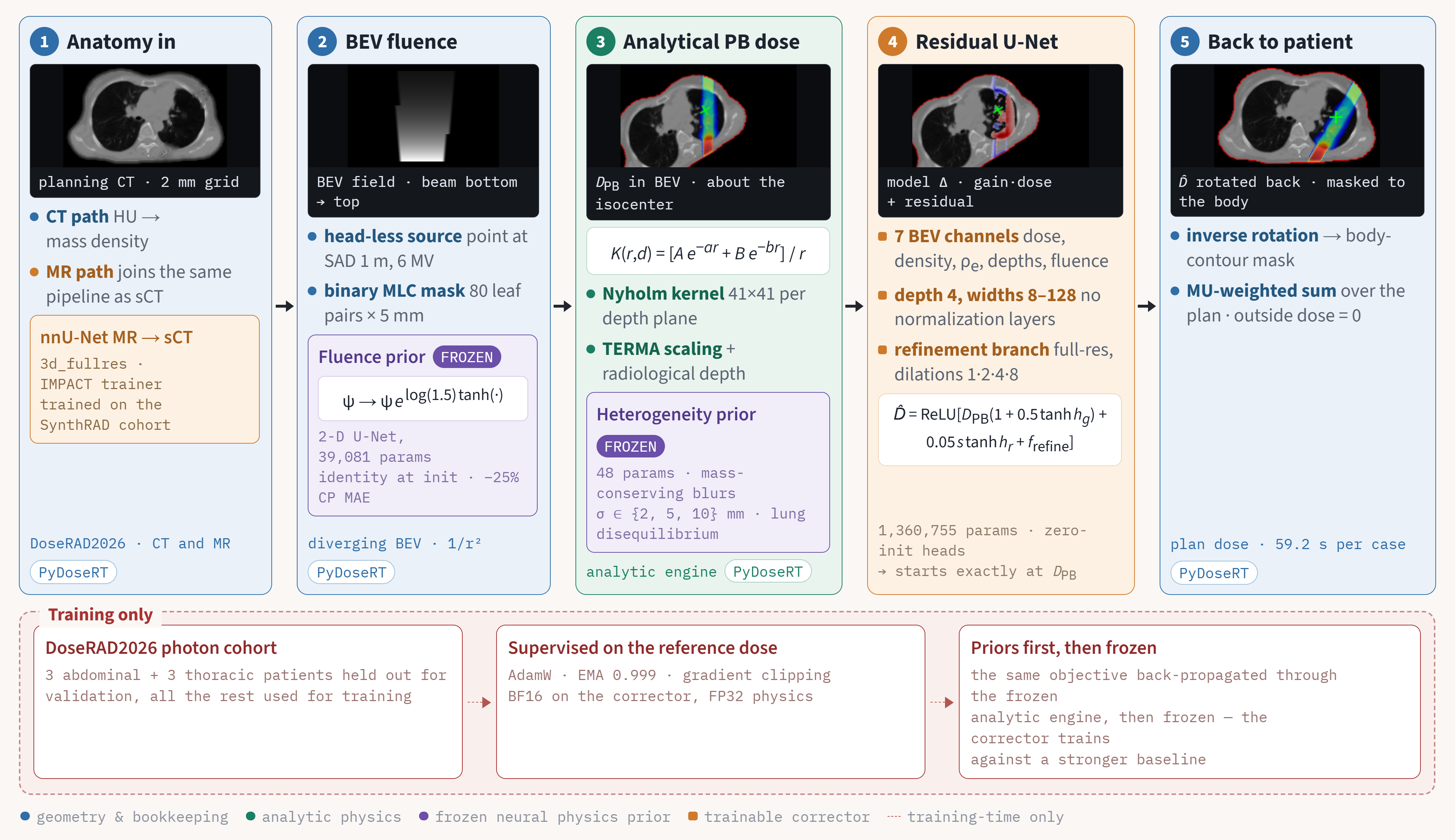}
\caption{PyDoseRT Photon workflow. The aperture fluence of each control point
is corrected by a frozen 2-D prior, projected into the BEV frame, and
convolved by the calibrated pencil-beam engine with TERMA source scaling; a
frozen 48-parameter prior redistributes dose laterally, and a compact 3-D
U-Net with a sequential refinement branch applies the bounded final
correction before rotation back to the patient grid. All learned stages are
zero-initialized and trained with the same peak-normalized per-CP loss.}
\label{fig:workflow}
\end{figure}

We used the DoseRAD2026 photon dataset~\cite{ref_doserad}: 69 training
patients (36 thoracic 1THB, 33 abdominal 1ABB) and 6 held-out validation
patients picked randomly (1ABB045/123/169, 1THB043/119/218) used for model selection. Each
patient provides a planning CT, a paired MR, 540 segment-dose samples constructed as three 180-segment arcs ($2^\circ$) gantry steps using three superior–inferior isocenter positions and one original or MLC-perturbed aperture per gantry angle with a
Geant4 MC reference dose each. All volumes share a $2\,\mathrm{mm}$ isotropic
grid. The learning target is the provided MC dose; inputs were the CT, from
which mass density is derived with the challenge HU--density table, and the
plan geometry (isocenter, gantry angle, 80 integer-millimetre MLC leaf pairs
of $5\,\mathrm{mm}$). The synthetic-CT (sCT) model was trained separately using all 75 paired MR/CT volumes. No additional patient data were used. Figure~\ref{fig:workflow} summarizes the pipeline.

\subsection{Model}
\label{sec:model}
The method has a three-stage, coarse-to-fine correction strategy, progressing from explicit physical modelling to increasingly flexible learned correction. First, the analytic
engine generates the baseline dose distribution. Second, two compact neural physics priors inside correct the incident fluence and lateral energy transport and were frozen after separate training. Third, a volumetric network corrects the remaining discrepancy from the MC reference. Eventually all model weights are frozen for evaluation.

\paragraph{Beam model.}
The challenge MC has no linac head: photons were rejection-sampled through a
binary $1\,\mathrm{mm}$ MLC mask and focused to a point at SAD
$1000\,\mathrm{mm}$, giving a geometrically sharp field edge and an
unflattened 6-MV spectrum. The engine therefore disables head scatter,
output factors, off-axis profiles, and MLC transmission. Leaf edges are
anti-aliased by fractional pixel coverage and blurred with a
$1.5\,\mathrm{mm}$ FWHM Gaussian, absorbing the aperture quantization and
grid binning rather than modelling a source. The fluence is projected into a
diverging BEV volume with the inverse-square factor per depth plane.

\paragraph{Pencil-beam dose engine.}
The lateral kernel is the analytical beam-quality-parameterized double
exponential of Nyholm et al.~\cite{ref_nyholm}, whose depth functions are
polynomials in $\mathrm{TPR}_{20,10}$; it is generated per depth plane at run
time ($41\times41$ voxels, $82\,\mathrm{mm}$) and applied as one grouped 2-D
convolution per (beam, depth). Kernels were indexed by a per-plane reference
radiological depth, the fluence-weighted mean of the divergent radiological
depth over the field's in-patient voxels (\emph{tile-mean} mode, robust to
sloped entrance surfaces); a residual per-voxel factor
$\exp[-\mu_\mathrm{eff}(d_\mathrm{rad}-d_\mathrm{ref})]$, clamped to
$[0.3,3]$ with $\mu_\mathrm{eff}=0.04\,\mathrm{cm}^{-1}$, restores the local
depth dependence. Before convolution, the published \emph{TERMA source
scaling model}~\cite{ref_laakkonen}  multiplies the local source term by an analytic factor depending on relative density and effective field size. Dose is scaled by the single fitted output
constant ($\mathrm{TPR}_{20,10}=0.646$ from the depth slope of
$\log(D_\mathrm{MC}/D_\mathrm{PB})$ in homogeneous abdominal soft tissue;
output $=6\,\mathrm{MV}\times$ the $L_1$-optimal scale), multiplied by the
MU weight, rotated into the patient frame by bilinear resampling, masked by
the body contour, and summed over beams. All physics runs on a bit-identical
BEV work crop ($3.6\times$ fewer voxels). A multilattice decomposition into
up to $5\times5$ equal-fluence tiles and a hand-fitted lateral-scatter
heuristic were both retired: the single tile is equal in accuracy at
$9$--$25\times$ less convolution, and the frozen priors below replace the
heuristic.

\paragraph{Frozen neural physics priors.}
Two small modules were trained by propagating the training loss
(Sect.~\ref{sec:training}) through the frozen engine, then frozen and
embedded in it. Both were identity at initialization and bounded.
(i)~\emph{Fluence prior}: a 2-D U-Net (39{,}081 parameters, base width 8,
two poolings) maps the analytic fluence map to a bounded log-gain,
$\psi\!\rightarrow\!\psi\,e^{\log(1.5)\tanh(\cdot)}$, before projection ---
absorbing whatever the binary-aperture rejection sampling and the sharp
penumbra model miss at the source plane. Trained for one epoch over the full
training cohort. (ii)~\emph{Lateral heterogeneity prior}: 48 parameters that,
after TERMA, choose a bounded signed fraction ($0.4\tanh$, with a soft density gate) of the local dose and redistribute it through a softmax mixture of
three fixed mass-conserving operators
$\mathrm{blur}_\sigma(\cdot)-(\cdot)$, $\sigma\in\{2,5,10\}\,\mathrm{mm}$,
conditioned on normalized dose, fluence, density, cumulative density path,
and the two lateral density gradients---a learned, spatially adaptive
replacement for lateral electronic-disequilibrium corrections in lung,
trained for one epoch on top of the frozen fluence prior.

\paragraph{Correction network.}
A 3-D U-Net (1{,}360{,}755 trainable parameters) operates per CP in BEV
space on the cropped volume. Its seven input channels are: corrected PB dose in
physical units, peak-normalized corrected PB dose, mass density, relative electron
density, radiological depth ($\int\rho_e\,\mathrm{d}l/100\,\mathrm{mm}$),
geometric depth from the isocenter, and the peak-normalized projected
fluence (kept continuous so sub-voxel aperture edges survive). Depth 4 with
widths 8--128, two $3^3$ convolutions per stage, ReLU, \emph{no
normalization layers} (masked-group, group, and instance normalization were
indistinguishable in accuracy and $\approx30\%$ slower), max-pool
down-sampling and trilinear up-sampling. A full-resolution dilated
refinement branch (hidden width 16, dilations 1, 2, 4, 8) runs
\emph{sequentially}: it receives the input stack plus the trunk's correction
as an extra channel and predicts a residual on top of it, rather than
independently estimating the same quantity (432 extra parameters, the best
of both coupling modes). Two $1\times1\times1$ heads produce the correction
\begin{equation}
\hat D=\mathrm{ReLU}\!\big[D_\mathrm{PB}\big(1+0.5\tanh h_g\big)
+0.2\,s\,\tanh h_r + f_\mathrm{refine}\big],
\label{eq:residual}
\end{equation}
where $s$ is the per-CP PB peak: a multiplicative gain bounded to $\pm50\%$
and an additive residual bounded to $\pm20\%$ of the peak. All heads are
zero-initialized, so the network begins exactly at the engine output. No
externally pre-trained weights were used. The seven-channel stack was fixed by a leave-one-out ablation: six
further candidate channels together contributed only $\approx4\%$ of the
model's own error and were dropped.

\paragraph{MR-to-CT conversion.}
Training used
$48\times192\times192$ patches at $3\times1\times1\,\mathrm{mm}^3$ spacing,
batch size 2, MR Z-score normalization (input), and global CT-intensity normalization (output).
The regression objective used a deep-supervised MAE equivalent to the nnUNets default loss. The fold-all run was trained for 1000 epochs with Nesterov SGD
(initial learning rate $0.01$, momentum $0.99$, weight decay $3\times10^{-5}$)
and polynomial learning-rate decay, without mirroring or geometric
augmentation. At inference, overlapping $48\times192\times192$ windows are
combined with Gaussian weighting; the resulting sCT is passed unchanged to the CT dose pipeline.

\subsection{Training}
\label{sec:training}
The corrector was trained on CT and evaluated in patient space on the
body-masked prediction. Let $t$ be the MC reference of one CP, $p_n=\max t$
its peak, and $e=|\hat D-t|/p_n$. The loss is
\begin{equation}
 \mathcal L=\big\langle\langle e\rangle_{t>0}\big\rangle_n
 +0.15\,\big\langle\langle e\rangle_{t\ge0.1p_n}\big\rangle_n,
 \label{eq:loss}
\end{equation}
averaged over valid CPs $n$ (closed-MLC CPs were dropped). The support term
scores every voxel with reference dose and the second term adds weight on
the high-dose region; the terms were additive, so neither can be traded away.
The same loss trains both physics priors.

Every CP of every training patient is used in each epoch. One engine forward
processes three CPs of the same patient and beam; eight such forwards are
accumulated per optimizer step (effective batch 24 CPs). AdamW (weight decay
$10^{-5}$), gradient-norm clipping at 1.0, EMA (decay 0.999) and epsilon at $10^{-10}$ used for
validation and checkpointing with a mean squared error loss, and BF16 autocast around the corrector only
(the physics stays FP32); no dropout or geometric augmentation. Training
proceeds at a constant $5\times10^{-4}$ learning rate. The anatomy-agnostic corrector is trained on all
69 patients and validated on 3 thoracic and 3 abdominal patients, on both CT
and sCT. Checkpoints were selected on validation Level-1 CP MAE with Level-2
plan gamma tracked alongside; the final version has reached $0.0094$.

For the separately ranked MRI track, the same model was used for dose correction.

\paragraph{Inference.}
One checkpoint serves both anatomies; the container loads per-modality
checkpoints for the CT and MRI tracks. The corrector and the sCT network run
in FP16 ($1.5\times$ faster per dose map at a $0.01$ percentage-point gamma
cost; $2.7\times$ faster sCT at $0.17\,\mathrm{HU}$ body deviation), whereas
the physics engine stays FP32. The engine is built once per
beam and reused across its CPs; outputs were thresholded by the per-CP
challenge cutoff, restored to the full grid, and written as compressed
stacks. The longitudinal crop margin of $60\,\mathrm{mm}$ was used, measured to buy $2$ gamma percentage points. During the submission we have used a batch size of 8.

\subsection{Evaluation}
\label{sec:eval}
We used the official DoseRAD2026 metrics with the vendored evaluator: Level~1
per-CP masked MAE (reference above 10\% of that CP's maximum, normalized by
it) and IDD curve distance (accumulated dose versus geometric depth,
normalized RMS), and Level~2 stratified plan MAE, local gamma pass rate
(1\%/1\,mm, 10\% threshold). The final
ranking averages the per-metric leaderboard positions with runtime counted
twice, so runtime carries a large portion of the ranking---more than any accuracy
metric---with a hard exclusion above $181\,\mathrm{s}$ per case. Local
validation therefore reported the full metric set per modality and anatomy,
estimating mean position rather than accuracy alone.

\section{Results}
\subsection{Validation set (6 held-out patients)}
Engine only, the fluence prior reduces validation of the CT task CP MAE from $0.0489$ to $0.0375$ and the heterogeneity prior reduces it to $0.0350$, acting mostly on thorax ($-10.5\%$) (Sect.~\ref{sec:model}). The corrector then reduces the remaining CP error to $0.0086$. Table~\ref{tab:test} reports the last local evaluation of our model submitted for the final tests.

\begin{table}[H]
\centering
\caption{Local evaluation metrics for the final submission evaluated on the 6 validation patients.}
\label{tab:test}
\small
\begin{tabular}{lccccc}
\toprule
Track & CP MAE & IDD & Plan MAE & Gamma (\%) & Runtime (s)\\
\midrule
CT & $0.0086$ & $0.0011$ & $0.0024$ & $99.12\%$ & $46\,\mathrm{s}$ \\
MRI & $0.0098$ & $0.0013$ & $0.0047$ & $96.95\%$ & $49\,\mathrm{s}$ \\
\bottomrule
\end{tabular}
\end{table}

\section{Discussion}
\label{sec:discussion}
The pipeline expresses one idea at three scales: learnable capacity is
placed where the physics says the analytic model is wrong, in the smallest
module that can express the correction, before falling back to a generic
volumetric network. Both priors were frozen, so the volumetric corrector
trains against a stable, stronger baseline and its bounded heads only need
to close what has no simple physical address; zero initialization at every
stage keeps the analytical solution as the fallback.

The second organizing principle is the ranking itself. With runtime counted
twice, a saved second is worth roughly twice a comparable accuracy gain, and
several accuracy-neutral choices followed directly: the single-tile
lattice, the bit-identical BEV work crop, the normalization-free
corrector ($\approx30\%$ faster at equal accuracy), and FP16 serving.
Conversely, capacity is nearly free---the corrector's cost is spatial
resolution, not width---so the accuracy effort concentrates on the loss and
the low-dose stratum, where the plan-level metrics fail: on the worst
validation patient, $\approx90\%$ of gamma failures lie in the 10--30\%
band. Additionally, the IDD, MAE and gamma pass rate metrics were often contradictory, at each design decision we were aiming to keep the metrics balanced, instead of favoring one metric over another.

Limitations include only six patients for local model selection, tuning and
reporting on the same validation cohort, as our local evaluation often did not reflect the preliminary leaderboard scores. An additional limitation is that by design, the MRI results will depend on the sCT quality.

\subsubsection{Author contributions (CRediT).} \textit{Attila Simk\'o} -- Conceptualization, Data curation, Formal analysis, Investigation, Methodology, Resources, Software, Validation, Visualization, Writing – original draft. \textit{Lukas Zimmermann} -- Investigation, Software, Writing – review \& editing. \textit{Hermann Fuchs} -- Resources, Writing – review \& editing. \textit{Gerd Heilemann} -- Project administration, Writing – review \& editing.

\subsubsection{Code availability.}
The photon dose engine builds on the open-source PyDoseRT
package.\footnote{\url{https://github.com/UMU-DDI/PyDoseRT}} The challenge container, training code, evaluation scripts, and model definitions will be available in a separate repository.

\begin{credits}
\subsubsection{\ackname}
Computational resources were provided by the National Academic Infrastructure for Supercomputing in Sweden (NAISS), funded by the Swedish Research Council through grant agreement no.~2025-5504.
\end{credits}


\begin{thebibliography}{8}

\bibitem{ref_ahnesjo}
Ahnesj\"o, A., Saxner, M., Trepp, A.: A pencil beam model for photon dose
calculation. Med. Phys. \textbf{19}, 263--273 (1992)

\bibitem{ref_pydose}
Simk\'o, A., Kronsteiner, M., Glatzer, S., et al.: A physics-informed,
plug-and-play dose engine for gradient-based radiotherapy treatment planning.
arXiv:2512.18863 (2025)

\bibitem{ref_doserad}
Xiao, F., Delopoulos, N., Wahl, N., et al.: DoseRAD2026 Challenge dataset:
AI accelerated photon and proton dose calculation for radiotherapy.
arXiv:2604.12778 (2026)

\bibitem{ref_nyholm}
Nyholm, T., Olofsson, J., Ahnesj\"o, A., Karlsson, M.: Photon pencil kernel
parameterisation based on beam quality index. Radiother. Oncol.
\textbf{78}, 347--351 (2006)

\bibitem{ref_laakkonen}
Linda Laakkonen and Zheyong Fan and Ari Harju: Technical note: TERMA scaling as an effective heterogeneity correction model for convolution-based external-beam photon dose calculations. Medical Physics. \textbf{50} 3191--3198 (2023)

\bibitem{ref_nnunet}
Isensee, F., Jaeger, P.F., Kohl, S.A.A., Petersen, J., Maier-Hein, K.H.:
nnU-Net: a self-configuring method for deep learning-based biomedical image
segmentation. Nat. Methods \textbf{18}, 203--211 (2021)

\end{thebibliography}
\end{document}